\documentclass[
aps,prd,
12pt,
nopreprintnumbers,
showpacs,
eqsecnum,
nofootinbib
]{revtex4-1}

\usepackage{graphicx}
\usepackage{amssymb}

\begin{document}

\title{Classical and quantum bicosmology with noncommutativity}
\author{Nahomi Kan}\email[]{kan@gifu-nct.ac.jp}
\affiliation{National Institute of Technology, Gifu College,
Motosu-shi, Gifu 501-0495, Japan}
\author{Takuma Aoyama}\email[]{b014vbv@yamaguchi-u.ac.jp}
\affiliation{
Graduate School of Sciences and Technology for Innovation, Yamaguchi
University, Yamaguchi-shi, Yamaguchi 753--8512, Japan}
\author{Kiyoshi Shiraishi}\email[]{shiraish@yamaguchi-u.ac.jp}
\affiliation{
Graduate School of Sciences and Technology for Innovation, Yamaguchi
University, Yamaguchi-shi, Yamaguchi 753--8512, Japan}
\date{\today}

\begin{abstract}
Recently, Falomir, Gamboa, M\'endez, Gondolo and
Maldonado proposed a bicosmology scenario \cite{FGMG,FGGM,FGM,MM} for solving some
cosmological problems related to inflation, dark matter, and thermal history of
the universe. Their plan is to introduce noncommutativity into the momentum space
of the two scale factors. In the present paper, we revisit their model and first
consider exact classical solutions in the model with constant noncommutativity
between dynamical variables and between canonical momenta.
We also hypothesize that the noncommutativity appears when the
scale factors are small, and show the behavior of the classical solution
 in that case with momentum-space noncommutativity.
Finally, we write down the Wheeler--DeWitt equation in that case and
examine the behavior of the solution.
\end{abstract}



\maketitle

\section{Introduction}
\label{sec1}
The bicosmology scenario has been proposed by Falomir, Gamboa, M\'endez, Gondolo
and Maldonado \cite{FGMG,FGGM,FGM,MM}.
In their scenario, two scale factors are introduced to represent two causally
separated regions of a universe, or even two different universes in a multiverse
scenario \cite{MM}.
From a different point of view, it can be regarded as a kind of
cosmological models in bigravity theory (for example, see Refs.~\cite{AM,SS,DM}).
However, instead of the mass term by mixing of two scale factors in bigravity,
mixing is guaranteed to be due to the noncommutativity in the
Hamiltonian formulation of the classical bicosmology.

Noncommutative cosmology has been studied by many authors
\cite{COR1,BP,PM,PO,AAOSS,GSS1,GSS2,GSS3,MOS1,MOS2,OMSS,SPOA,BBDP1,BBDP2,BBDP3,
MMP,OQ,RM1,RM2,RZJM,JCM,RSFMM,RMM,KKST2,JRN,COR2,Rasouli}%
\footnote{There are a huge number of papers on noncommutative cosmology,
and it is difficult to cite all of them. We apologize if we have overlooked any
important papers.},  not only classically
but also quantum mechanically. Thus, we intend to study various aspects of
classical and quantum bicosmology.

In the present paper, we revisit the model of bicosmology and examine classical
and quantum cosmological solutions. Although the main analyses in the later
sections will be illustrated when canonical momenta are noncommutative, we also
consider noncommutativity in dynamical variables in the minisuperspace at first.

Next, we observe the case where the parameter representing the noncommutativity
is not constant.
The most natural idea is that noncommutativity has little effect on the
cosmological evolution when the universe is large, but is highly effective at the
very early stage of the universe, when the universe is small.
We also write down the Wheeler--DeWitt (WDW) equation for that case and
study the behavior of the solution.

Sec.~\ref{sec2} describes setting up the model for bicosmology
and obtaining classical solutions for the commutative case.
In Sec.~\ref{sec3}, we study classical bicosmology with noncommutativity in the
phase space. In Sec.~\ref{sec4}, we define the nonconstant
noncommutativity in momentum space and investigate classical solutions in the
system. Quantum bicosmology by the WDW equation in the system is studied in
Sec.~\ref{sec5}. Finally, conclusion and discussion are given in the last section. 

\section{commutative bicosmology model}
\label{sec2}

We consider the following action consisting only of the Einstein--Hilbert term and
the cosmological term:
\begin{equation}
S\propto\int d^Dx\sqrt{-g}\Bigl[R-2\Lambda\Bigr]\,,
\label{SS}
\end{equation}
where  $R$ is the Ricci scalar derived from the metric $g_{\mu\nu}$
($\mu,\nu=0, 1,\ldots,D-1$), and $g$ is the determinant of $g_{\mu\nu}$. 
We also assumed that the cosmological constant $\Lambda$ is a positive constant.
As a first step in an analysis, it is very important to investigate cases where
such toy models have exact solutions.
It will also be useful for consideration on numerical analyses of sophisticated
models with further variations in the future.  

First, in this section we will review the commutative version of the bicosmology,
which we will use later for comparison with the noncommutative case.

We take an ansatz for the metric tensor as
\begin{equation}
ds^2=g_{\mu\nu}dx^\mu dx^\nu=-N^2dt^2+a^{2}(t)d\mathbf{x}^2\,,
\end{equation}
where $t=x^0$, $d\mathbf{x}^2\equiv\sum_{i=1}^{D-1}(dx^{i})^2$, and
$N$ stands for the lapse function.
Then, one can derive the following Lagrangian for the scale factor from the action
(\ref{SS}):
\begin{equation}
L=-\frac{1}{2N}a^{{D-3}}\dot{a}^2-\frac{N}{(D-1)(D-2)}\Lambda a^{D-1}\,,
\end{equation}
where the dot indicates the derivative with respect to time $t$ and
total derivatives have been dropped.
Now, we switch the dynamical variable $a$ to $x$, which is defined as
\begin{equation}
x\equiv\frac{2}{D-1} a^{\frac{D-1}{2}}\,.
\end{equation}
Then, the Lagrangian takes the form
\begin{equation}
L=-\frac{1}{2N}\dot{x}^2-\frac{N}{2}\gamma^2 x^2\,,
\end{equation}
where
\begin{equation}
\gamma^2=\frac{D-1}{2(D-2)}\Lambda\,.
\end{equation}
Note that this Lagrangian is just that of an inverted harmonic oscillator.

In the bicosmology scenario \cite{FGMG,FGGM,FGM,MM}, the second scale factor $b$,
which represents the causally disconnected region, is introduced.
Here, we adopt the case that the Lagrangian for the second scale factor $b$ is
just a copy of that for the first scale factor $a$, though the authors of
Refs.~\cite{FGMG,FGGM,FGM,MM} considered generally different values for
cosmological constants. Therefore, the total Lagrangian takes the form of the
two-dimensional inverted harmonic oscillator:
\begin{equation}
L=-\frac{1}{2N}(\dot{x}^2+\dot{y}^2)-\frac{N}{2}\gamma^2 (x^2+y^2)\,,
\end{equation}
where
\begin{equation}
y=\frac{2}{D-1} b^{\frac{D-1}{2}}\,.
\end{equation}
To construct the Hamiltonian, we have only to follow the standard method.
The conjugate momenta are found to be
\begin{equation}
\pi_x\equiv\frac{\partial
L}{\partial \dot{x}}=-\frac{\dot{x}}{N}\,,\qquad
\pi_y\equiv\frac{\partial
L}{\partial \dot{y}}=-\frac{\dot{y}}{N}\,,
\end{equation}
and the Hamiltonian is finally obtained as
\begin{equation}
H=N\left[-\frac{1}{2}(\pi_x^2+\pi_y^2)+\frac{1}{2}\gamma^2 (x^2+y^2)\right]\,,
\end{equation}
which is the Hamiltonian for the two-dimensional inverted harmonic oscillator
(up to the lapse function).

The Hamiltonian constraint is assigned from $\frac{\partial H}{\partial N}=0$.
We thus set $N=1$ hereafter.



The Poisson brackets are placed as usual:
\begin{equation}
\{x, \pi_x\}=\{y, \pi_y\}=1\,,
\label{comPoi}
\end{equation}
and others are zero.
Then, the usual Hamilton's equations are obtained as
\begin{eqnarray}
& &\dot{x}=\{x, H\}=\frac{\partial H}{\partial\pi_x}=-\pi_x\,,\quad
\dot{y}=\{y, H\}=\frac{\partial H}{\partial\pi_y}=-\pi_y\,,\nonumber \\
& &\dot{\pi}_x=\{\pi_x, H\}=-\frac{\partial H}{\partial x}=-\gamma^2x\,,\quad
\dot{\pi}_y=\{\pi_y, H\}=-\frac{\partial H}{\partial y}=-\gamma^2y\,,
\end{eqnarray}
and subsequently,
we can derive the equations of motion as
\begin{equation}
\ddot{x}-\gamma^2 x=0\,,\quad
\ddot{y}-\gamma^2 y=0\,.
\label{eom}
\end{equation}
The solution of the equations is
\begin{eqnarray}
x(t)&=&C_1\cos\varphi_0\exp[\gamma t]+C_2\sin\varphi_0\exp[-\gamma t]\,, \\ 
y(t)&=&C_1\sin\varphi_0\exp[\gamma t]-C_2\cos\varphi_0\exp[-\gamma t]\,, 
\end{eqnarray}
where the arrangement of the constant coefficients $C_1$, $C_2$, and the
constant phase angle $\varphi_0$ have been chosen as the Hamiltonian constraint
$H=0$ holds.
These exponential behaviors are expected from the inverse harmonic oscillator
Hamiltonian.

With the preparatory calculations complete, in the next section we consider
classical noncommutative bicosmology.

\section{Classical noncommutative bicosmology}
\label{sec3}

The authors of bicosmology \cite{FGMG,FGGM,FGM,MM}
introduced noncommutativity into the momentum space of the system.
We first consider noncommutative dynamical variables as well.

We presume that the corresponding noncommutative system is defined
by replacement%
\footnote{Note that the usage of letters may be slightly different from the
previous work in bicosmology \cite{FGMG,FGGM,FGM,MM}.}
\begin{equation}
x\rightarrow X\,,\quad \pi_x\rightarrow\Pi_X\,,\quad
y\rightarrow Y\,,\quad \pi_y\rightarrow\Pi_Y\,,
\end{equation}
and the Hamiltonian takes the form of the inverted harmonic oscillator,
\begin{equation}
H_{NC}=-\frac{1}{2}(\Pi_X^2+\Pi_Y^2)+\frac{1}{2}\gamma^2 (X^2+Y^2)\,.
\label{ncH}
\end{equation}
Here, we assume the noncommutative phase space in the minisuperspace,
which is realized by the deformed Poisson brackets
\begin{equation}
\{X, \Pi_X\}=\{Y, \Pi_Y\}=1\,,\quad \{X, Y\}=\Theta\,,\quad
\{\Pi_X, \Pi_Y\}=B\,,
\label{ncPoi}
\end{equation}
and others are zero. We assume the parameters which
represent noncommutativities, $\Theta$ and $B$, are constant for this time.

Now, the canonical equations can be obtained as
\begin{eqnarray}
& &\dot{X}=\{X, H_{NC}\}=-\Pi_X+\gamma^2\Theta Y\,,\quad
\dot{Y}=\{Y, H_{NC}\}=-\Pi_Y-\gamma^2\Theta X\,,\nonumber \\
& &\dot{\Pi}_X=\{\Pi_X, H_{NC}\}=-\gamma^2X-B\Pi_Y\,,\quad
\dot{\Pi}_Y=\{\Pi_Y, H_{NC}\}=-\gamma^2Y+B\Pi_X\,,
\label{nch}
\end{eqnarray}
and consequently, the equations of motion for $X$ and $Y$ can be revealed as
\begin{equation}
\ddot{X}-(\gamma^2\Theta-B)\dot{Y}-\gamma^2(1-B\Theta)X=0\,,\quad
\ddot{Y}+(\gamma^2\Theta-B)\dot{X}-\gamma^2(1-B\Theta)Y=0\,.
\end{equation}
These equations can be easily solved, especially if we considered a combination
\begin{equation}
\ddot{Z}+i(\gamma^2\Theta-B)\dot{Z}-\gamma^2(1-B\Theta)Z=0\,,
\end{equation}
where $Z\equiv X+iY$.

The general solutions which satisfy the Hamiltonian constraint $H_{NC}=0$ are
exhibited as follows.

\noindent
$\bullet$ if $|B+\gamma^2\Theta|<2\gamma$
\begin{eqnarray}
X(t)&=&C_1\exp\Biggl[\frac{\sqrt{4\gamma^2-(\gamma^2\Theta+B)^2}}{2}t\Biggr]
\cos\Biggl[\frac{B-\gamma^2\Theta}{2}t+\varphi_1\Biggr]
\nonumber \\
&+&C_2\exp\Biggl[-\frac{\sqrt{4\gamma^2-(\gamma^2\Theta+B)^2}}{2}t\Biggr]
\sin\Bigl[\frac{B-\gamma^2\Theta}{2}t+\varphi_2\Bigr]\,, \\ 
Y(t)&=&C_1\exp\Biggl[\frac{\sqrt{4\gamma^2-(\gamma^2\Theta+B)^2}}{2}t\Biggr]
\sin\Biggl[\frac{B-\gamma^2\Theta}{2}t+\varphi_1\Biggr]
\nonumber \\
&-&C_2\exp\Biggl[-\frac{\sqrt{4\gamma^2-(\gamma^2\Theta+B)^2}}{2}t\Biggr]
\cos\Biggl[\frac{B-\gamma^2\Theta}{2}t+\varphi_2\Biggr]\,, 
\end{eqnarray}
where $C_1$, $C_2$, $\varphi_1$, and $\varphi_2$ are constants. The two phase
angles should satisfy
\begin{equation}
\sin(\varphi_2-\varphi_1)=\frac{B+\gamma^2\Theta}{2\gamma}\,,
\end{equation}
for the Hamiltonian constraint.

\noindent
$\bullet$ if $|B+\gamma^2\Theta|=2\gamma$
\begin{eqnarray}
X(t)&=&C
\cos\Biggl[\frac{B-\gamma^2\Theta}{2}t+\varphi_1\Biggr]
\nonumber \\
&+&C (B+\gamma^2\Theta)\sin(\varphi_1-\varphi_2)\cdot t
\cos\Biggl[\frac{B-\gamma^2\Theta}{2}t+\varphi_2\Biggr]\,,
\\  Y(t)&=&C
\sin\Biggl[\frac{B-\gamma^2\Theta}{2}t+\varphi_1\Biggr]
\nonumber \\
&+&C  (B+\gamma^2\Theta)\sin(\varphi_1-\varphi_2)\cdot t
\sin\Biggl[\frac{B-\gamma^2\Theta}{2}t+\varphi_2\Biggr]\,, 
\end{eqnarray}
where $C$, $\varphi_1$, and $\varphi_2$ are constants.

\noindent
$\bullet$ if $|B+\gamma^2\Theta|>2\gamma$
\begin{eqnarray}
X(t)&=&C e^{-\eta}
\cos\Biggl[\frac{B-\gamma^2\Theta+\sqrt{(B+\gamma^2\Theta)^2-4\gamma^2}}{2}t+
\varphi_1\Biggr]
\nonumber \\
&+&C e^{\eta}
\sin\Biggl[\frac{B-\gamma^2\Theta-\sqrt{(B+\gamma^2\Theta)^2-4\gamma^2}}{2}t+
\varphi_2\Biggr]\,,
\label{BX}
\\  Y(t)&=&C e^{-\eta}
\sin\Biggl[\frac{B-\gamma^2\Theta+\sqrt{(B+\gamma^2\Theta)^2-4\gamma^2}}{2}t+
\varphi_1\Biggr]
\nonumber \\
&-&C e^{\eta}
\cos\Biggl[\frac{B-\gamma^2\Theta-\sqrt{(B+\gamma^2\Theta)^2-4\gamma^2}}{2}t+
\varphi_2\Biggr]\,, 
\label{BY}
\end{eqnarray}
where $C$, $\varphi_1$, and $\varphi_2$ are constants. The constant $\eta$
is determined by
\begin{equation}
\cosh(2\eta)=\frac{|B+\gamma^2\Theta|}{2\gamma}\,.
\end{equation}

All these noncommutative classical solutions are always involve
trigonometric functions, with each scale factor going to zero at some point in
the past or the future.
The case with $B=\gamma^2\Theta$ is interesting because it has no slow
oscillatory behavior, but we naturally think that $B$ and $\gamma^2\Theta$ do not
take large values at present, and they are expected to be much less than
$H_0$ (the Hubble constant), so they do not
affect the current expansion of the universe.  Therefore, the
relation
$B\sim\gamma^2\Theta$ need not be taken too seriously. Interestingly,
however, it is possible that the parameters of the noncommutativity became
large only in the early stage of the small universe. We explore this possibility
 in the next section.


Incidentally, $X^2+Y^2$ in each case is found to be
\begin{eqnarray}
X^2+Y^2&=&
C_1^2\exp\Bigl[{\sqrt{4\gamma^2-(\gamma^2\Theta+B)^2}}t\Bigr]
+C_2^2\exp\Bigl[-{\sqrt{4\gamma^2-(\gamma^2\Theta+B)^2}}t\Bigr]\nonumber \\
& &+2C_1C_2\frac{B+\gamma^2\Theta}{2\gamma}\,,
\qquad\qquad\qquad\qquad\qquad\qquad (|B+\gamma^2\Theta|<2\gamma)
 \\
X^2+Y^2&=&
C^2\Bigl[1+2(B+\gamma^2\Theta)\sin(\varphi_1-\varphi_2)\cos(\varphi_1-\varphi_2)\cdot
t
\nonumber \\& &\qquad+
(B+\gamma^2\Theta)^2\sin^2(\varphi_1-\varphi_2)\cdot t^2\Bigr]\,,\qquad\qquad
(|B+\gamma^2\Theta|=2\gamma)
\\  
X^2+Y^2&=&
2C^2\Biggl[\frac{|B+\gamma^2\Theta|}{2\gamma}-
\sin[\sqrt{(B+\gamma^2\Theta)^2-4\gamma^2}t+\varphi_1-\varphi_2]\Biggr]\,.
\nonumber\\
& & \hspace{240pt}(|B+\gamma^2\Theta|>2\gamma)
\end{eqnarray}
It can be seen that $X^2+Y^2$ does not oscillate slowly, except for the rapid
oscillation for $|B+\gamma^2\Theta|>2\gamma$, so it can be a
measure of the average size of the two scale factors at least, but its
physical meaning should be discussed in later sections.


In the rest of this section, we will note the formulation of noncommutative
system using undeformed canonical variables.
A similar formulation is already known in quantum cosmology, the
technique required to find the WDW equation
\cite{COR1,BP,PM,PO,AAOSS,GSS1,GSS2,GSS3,MOS1,MOS2,OMSS,SPOA,BBDP1,BBDP2,BBDP3,
MMP,OQ,KKST2,JRN,COR2}.

For this, we set the variables as follows:
\begin{eqnarray}
& &X=\frac{1}{\sqrt{1+\frac{\beta\theta}{4}}}\Bigl(x-\frac{\theta}{2}\pi_y\Bigr)
\,,\quad
Y=\frac{1}{\sqrt{1+\frac{\beta\theta}{4}}}\Bigl(y+\frac{\theta}{2}\pi_x\Bigr)
\,,\nonumber \\
&
&\Pi_X=\frac{1}{\sqrt{1+\frac{\beta\theta}{4}}}\Bigl(\pi_x+\frac{\beta}{2}y\Bigr)
\,,\quad
\Pi_Y=\frac{1}{\sqrt{1+\frac{\beta\theta}{4}}}\Bigl(\pi_y-\frac{\beta}{2}x\Bigr)
\,,
\label{cncr}
\end{eqnarray}
where $x$, $\pi_x$, $y$, and $\pi_y$ obey the usual 
Poisson brackets (\ref{comPoi}).
If the constant parameters $\theta$ and $\beta$ are related to $\Theta$ and $B$ as
\begin{equation}
\Theta=\frac{\theta}{1+\frac{\beta\theta}{4}}\,,\quad
B=\frac{\beta}{1+\frac{\beta\theta}{4}}\,,
\end{equation}
one can find that the brackets for $X$, $\Pi_X$, $Y$, $\Pi_Y$ defined by
(\ref{cncr}) realize the noncommutative Poisson brackets
(\ref{ncPoi}).
Then, the Hamiltonian $H_{NC}$ is expressed as
\begin{equation}
H_{NC}=-
\frac{1-\frac{\gamma^2\theta^2}{4}}{2\left(1+\frac{\beta\theta}{4}\right)}
(\pi_x^2+\pi_y^2)
+\frac{\beta-\gamma^2\theta}{2\left(1+\frac{\beta\theta}{4}\right)}(x\pi_y-y\pi_x)
+\frac{\gamma^2-\frac{\beta^2}{4}}{2\left(1+\frac{\beta\theta}{4}\right)}
(x^2+y^2)\,.
\end{equation}

The equivalence of the equations motion can be checked easily.
Since one can see
\begin{eqnarray}
& &\dot{x}=\frac{\partial
H_{NC}}{\partial\pi_x}=-
\frac{1-\frac{\gamma^2\theta^2}{4}}{1+\frac{\beta\theta}{4}}\pi_x
-\frac{\beta-\gamma^2\theta}{2(1+\frac{\beta\theta}{4})}y\,,\nonumber \\
& &\dot{y}=\frac{\partial
H_{NC}}{\partial\pi_y}=-
\frac{1-\frac{\gamma^2\theta^2}{4}}{1+\frac{\beta\theta}{4}}\pi_y
+\frac{\beta-\gamma^2\theta}{2(1+\frac{\beta\theta}{4})}x\,,\nonumber
\\& &\dot{\pi}_x=-\frac{\partial
H_{NC}}{\partial x}=-
\frac{\gamma^2-\frac{\beta^2}{4}}{1+\frac{\beta\theta}{4}}x
-\frac{\beta-\gamma^2\theta}{2(1+\frac{\beta\theta}{4})}\pi_y\,,\nonumber \\
& &\dot{\pi}_y=-\frac{\partial
H_{NC}}{\partial y}=-
\frac{\gamma^2-\frac{\beta^2}{4}}{1+\frac{\beta\theta}{4}}y
+\frac{\beta-\gamma^2\theta}{2(1+\frac{\beta\theta}{4})}\pi_x\,,
\end{eqnarray}
one can finally find the identical equations with (\ref{nch}),
\begin{eqnarray}
& &\dot{X}=\frac{1}{\sqrt{1+\frac{\beta\theta}{4}}}
\Bigl(\dot{x}-\frac{\theta}{2}\dot{\pi}_y\Bigr)=-\Pi_X+\gamma^2\Theta Y
\,,\nonumber \\
& &\dot{Y}=\frac{1}{\sqrt{1+\frac{\beta\theta}{4}}}
\Bigl(\dot{y}+\frac{\theta}{2}\dot{\pi}_x\Bigr)=-\Pi_Y-\gamma^2\Theta X
\,,\nonumber \\
& &\dot{\Pi}_X=\frac{1}{\sqrt{1+\frac{\beta\theta}{4}}}
\Bigl(\dot{\pi}_x+\frac{\beta}{2}\dot{y}\Bigr)=-\gamma^2X-B\Pi_Y
\,,\nonumber \\
& &
\dot{\Pi}_Y=\frac{1}{\sqrt{1+\frac{\beta\theta}{4}}}
\Bigl(\dot{\pi}_y-\frac{\beta}{2}\dot{x}\Bigr)=-\gamma^2Y+B\Pi_X\,.
\end{eqnarray}
Then, the check is completed as mentioned above.

\section{noncommutativity in momenta for small scale factors: Classical system}
\label{sec4}

If the parameter $\theta=0$, (\ref{cncr}) is simplified as
\begin{equation}
X=x
\,,\quad
Y=y
\,,\quad\Pi_X=\pi_x+\frac{\beta}{2}y
\,,\quad
\Pi_Y=\pi_y-\frac{\beta}{2}x
\,.
\label{41}
\end{equation}
This is exactly what the authors of Refs.~\cite{FGMG,FGGM,FGM,MM} considered.
With the noncommutativity of the canonical momenta, this model has several
advantages over simplicity.
One advantage is that configuration variables $X$ and $Y$ are trivially
represented by $x$ and $y$. In particular, for the study of noncommutative
quantum cosmology in general, it is necessary to derive the Wigner function
\cite{Wigner,Case,WF} to analyze the solution of WDW equation for
$\{X,Y\}\ne 0$ (see Ref.~\cite{KKST2} for example). Even with the classical
model of cosmology, the analysis becomes 
tedious, at least in general cases. Models involving general noncommutativity
in phase spaces of the minisuperspaces are left for future work.

Another advantage of considering only the nontrivial Poisson bracket
$\{\Pi_X,\Pi_Y\}\ne 0$ is the analogy with a magnetic flux
\cite{FGMG,FGGM,FGM,MM}. In Ref.~\cite{FGGM}, the authors explicitly pointed to
the quantum Hall effect as a similar system.%
\footnote{The analogy is pointed out also by the recent paper \cite{GS}.}

Here, in the present paper, we propose nonconstant noncommutativity in the
minisuperspace.
As already mentioned, the noncommutativity is thought to have no effect
in the present epoch, but it may have a large effect in the early
universe when the scale factors are very small.

Now, suppose that the parameter $\beta$ in (\ref{41}) is a function of
$r=\sqrt{x^2+y^2}$. Accordingly, the Poisson bracket of $\Pi_X$ and $\Pi_Y$ becomes
\begin{equation}
\{\Pi_X, \Pi_Y\}=\beta(r)+\frac{r\beta'(r)}{2}\,,
\end{equation}
where the prime denotes the derivative with respect to $r$.
Thus, the relation between $\beta$ and $B$ is slightly modified if 
there are position dependence in the minisuperspace.

The Hamiltonian (\ref{ncH}) now takes the form
\begin{equation}
H_{NC}=-\frac{1}{2}
(\pi_x^2+\pi_y^2)
+\frac{1}{2}\beta(r)(x\pi_y-y\pi_x)+\frac{1}{2}
\left(\gamma^2-\frac{\beta^2(r)}{4}\right)
(x^2+y^2)\,.
\end{equation}
The canonical equations
\begin{eqnarray}
& &\dot{x}=\frac{\partial
H_{NC}}{\partial\pi_x}=-\pi_x
-\frac{\beta(r)}{2}y\,,\quad\dot{y}=\frac{\partial
H_{NC}}{\partial\pi_y}=-
\pi_y+\frac{\beta(r)}{2}x\,,\nonumber
\\& &\dot{\pi}_x=-\frac{\partial
H_{NC}}{\partial x}=-
\left(\gamma^2-\frac{\beta(r)(\beta(r)+r\beta'(r))}{4}\right)x
-\frac{\beta(r)}{2}\pi_y
-\frac{\beta'(r)x}{2r}(x\pi_y-y\pi_x)
\,,\nonumber \\
& &\dot{\pi}_y=-\frac{\partial
H_{NC}}{\partial y}=-
\left(\gamma^2-\frac{\beta(r)(\beta(r)+r\beta'(r))}{4}\right)y
+\frac{\beta(r)}{2}\pi_x
-\frac{\beta'(r)y}{2r}(x\pi_y-y\pi_x)
\,,
\end{eqnarray}
leads to the following equations of motion:
\begin{eqnarray}
& &\ddot{x}+\beta\dot{y}+\frac{\beta'}{2r}(x\dot{x}+y\dot{y})y-
\left(\gamma^2-\frac{r\beta\beta'}{4}\right)x+
\frac{\beta'}{2r}\left[x\dot{y}-y\dot{x}-\frac{\beta}{2}(x^2+y^2)\right]x=0\,,\\
& &\ddot{y}-\beta\dot{x}-\frac{\beta'}{2r}(x\dot{x}+y\dot{y})x-
\left(\gamma^2-\frac{r\beta\beta'}{4}\right)y+
\frac{\beta'}{2r}\left[x\dot{y}-y\dot{x}-\frac{\beta}{2}(x^2+y^2)\right]y=0\,.
\end{eqnarray}
By use of the polar coordinates
\begin{equation}
X=x=r\cos\phi\,,\quad Y=y=r\sin\phi\,,
\label{ct}
\end{equation}
the equations of motion reduce to
\begin{eqnarray}
& &\ddot{r}-r\dot{\phi}^2+\beta\, r\dot{\phi}-
\gamma^2 r+\frac{\beta'}{2}r^2\dot{\phi}\nonumber\\
&=&\ddot{r}-r\left(\dot{\phi}-\frac{\beta}{2}\right)^2-
\left(\gamma^2-\frac{\beta(\beta+r\beta')}{4}\right)r+\frac{\beta'}{2}r^2
\left(\dot{\phi}-\frac{\beta}{2}\right)=0\,,\\
& &r\ddot{\phi}+2\dot{r}\dot{\phi}-\beta\, \dot{r}
-\frac{\beta'}{2}r\dot{r}=\frac{1}{r}\frac{d}{dt}\left[r^2\left(\dot{\phi}-\frac{\beta}{2}\right)\right]=0\,.
\label{L}
\end{eqnarray}
On the other hand,
the Hamiltonian constraint reads
\begin{eqnarray}
H_{NC}&=&-\frac{1}{2}\dot{r}^2-\frac{1}{2}r^2\dot{\phi}^2
+\frac{1}{2}\gamma^2r^2\nonumber \\
&=&-\frac{1}{2}\dot{r}^2-\frac{1}{2}r^2\left(\dot{\phi}-\frac{\beta}{2}\right)^2
-\frac{\beta}{2}r^2\left(\dot{\phi}-\frac{\beta}{2}\right)
+\frac{1}{2}\left(\gamma^2-\frac{\beta^2}{4}\right)r^2=0\,.
\end{eqnarray}
The equation (\ref{L}) can be integrated and if we set
\begin{equation}
L=r^2\left(\dot{\phi}-\frac{\beta(r)}{2}\right)=\mbox{constant}\,,
\end{equation}
the constraint equation becomes
\begin{equation}
\dot{r}^2+\frac{L^2}{r^2}
+{L\beta(r)}
-\left(\gamma^2-\frac{\beta^2(r)}{4}\right)r^2=0\,,
\end{equation}
or, equivalently
\begin{equation}
\dot{r}^2=\gamma^2r^2-\left(\frac{L}{r}
+\frac{r\beta(r)}{2}\right)^2\,.
\label{eq2}
\end{equation}

Now, we find that the equations of motion can be integrated and have very simple
forms even if
$\beta$ is not constant as long as $\beta$ is an arbitrary function of $r$ only.

\begin{figure}[ht]
\centering
\includegraphics[width=3.1cm]{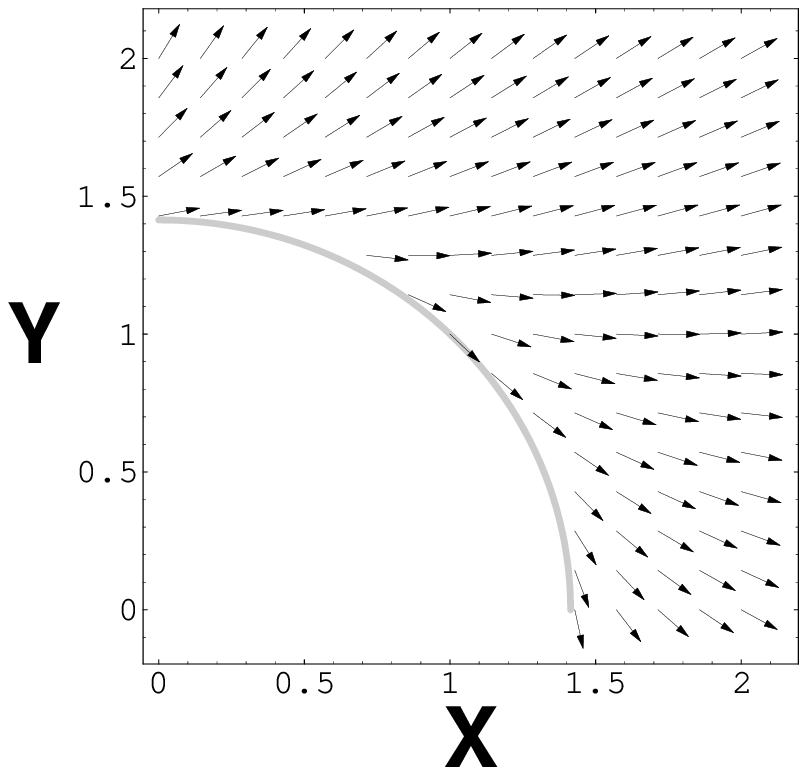}
\includegraphics[width=3.1cm]{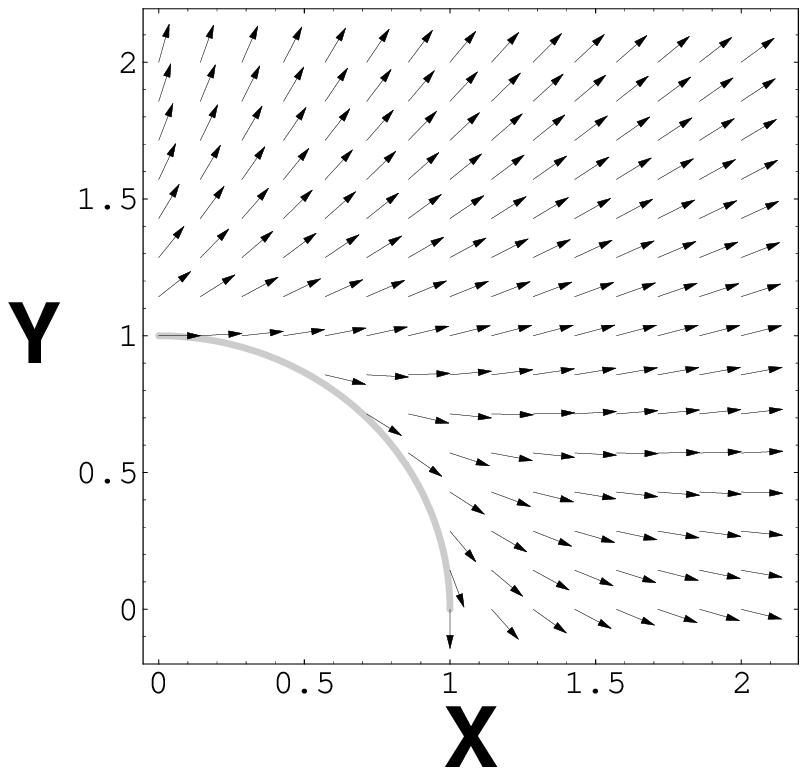}
\includegraphics[width=3.1cm]{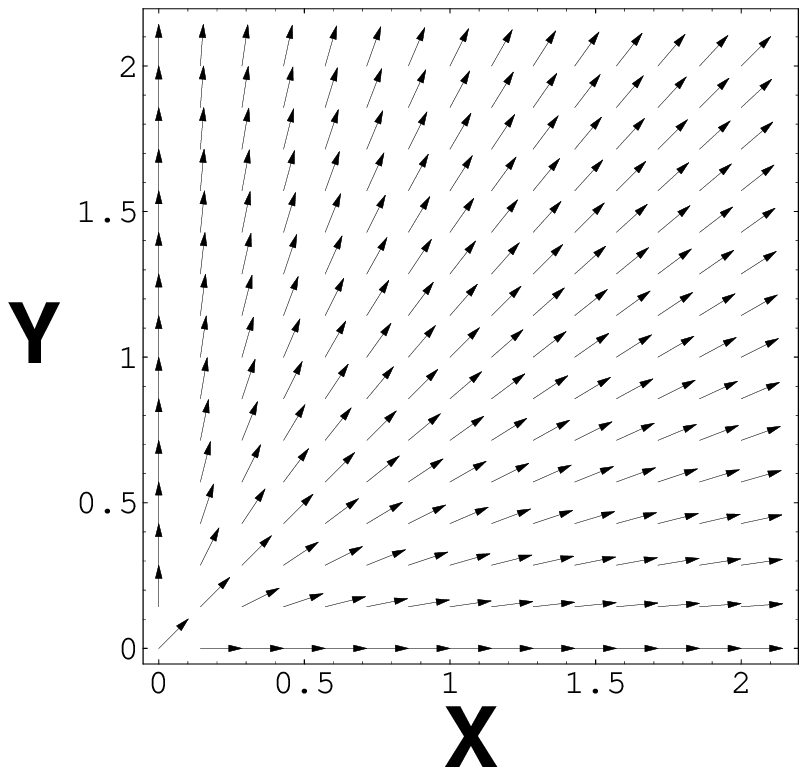}
\includegraphics[width=3.1cm]{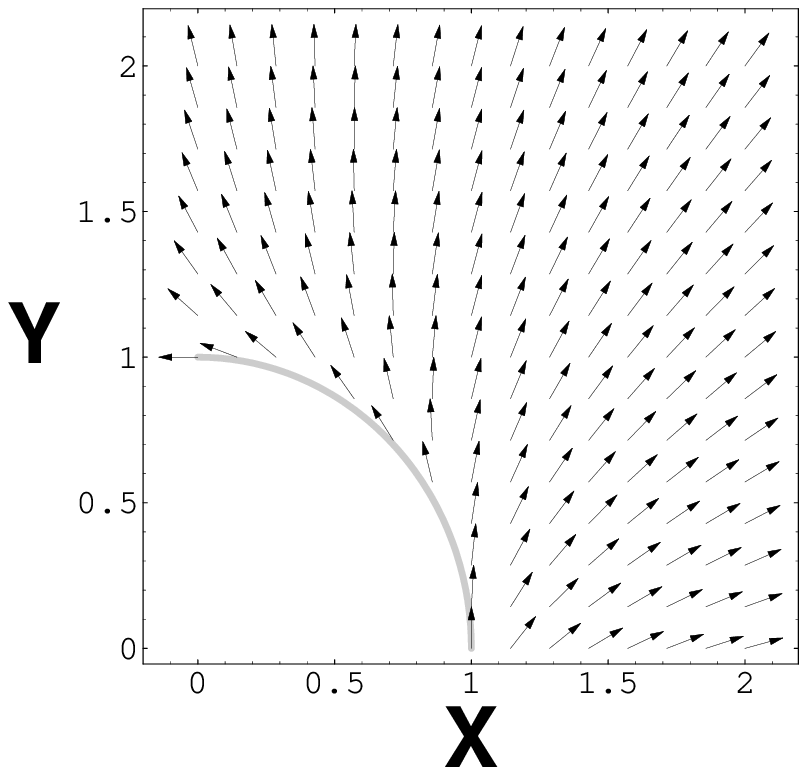}
\includegraphics[width=3.1cm]{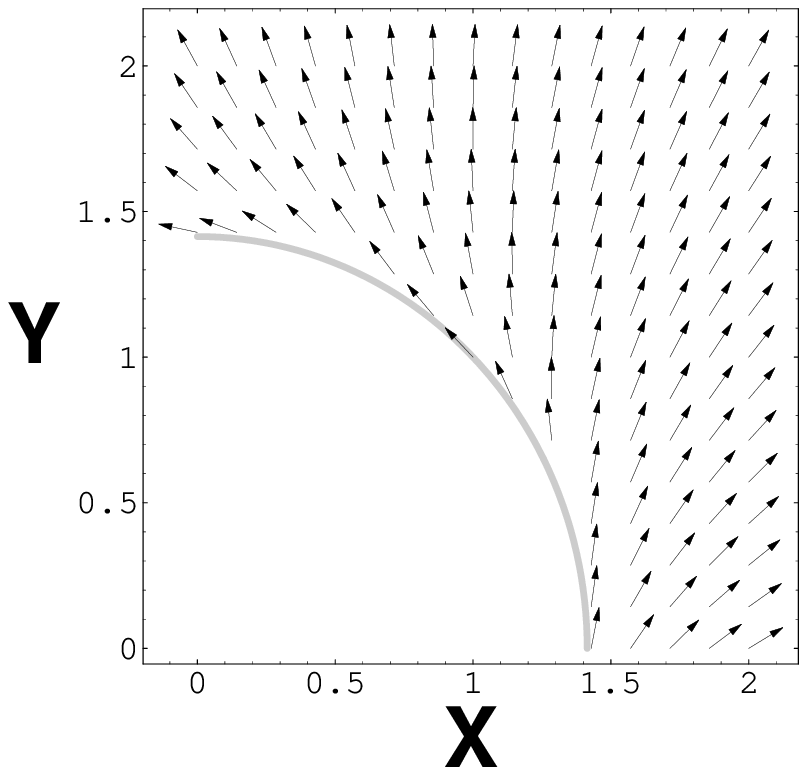}
\\
\includegraphics[width=3.1cm]{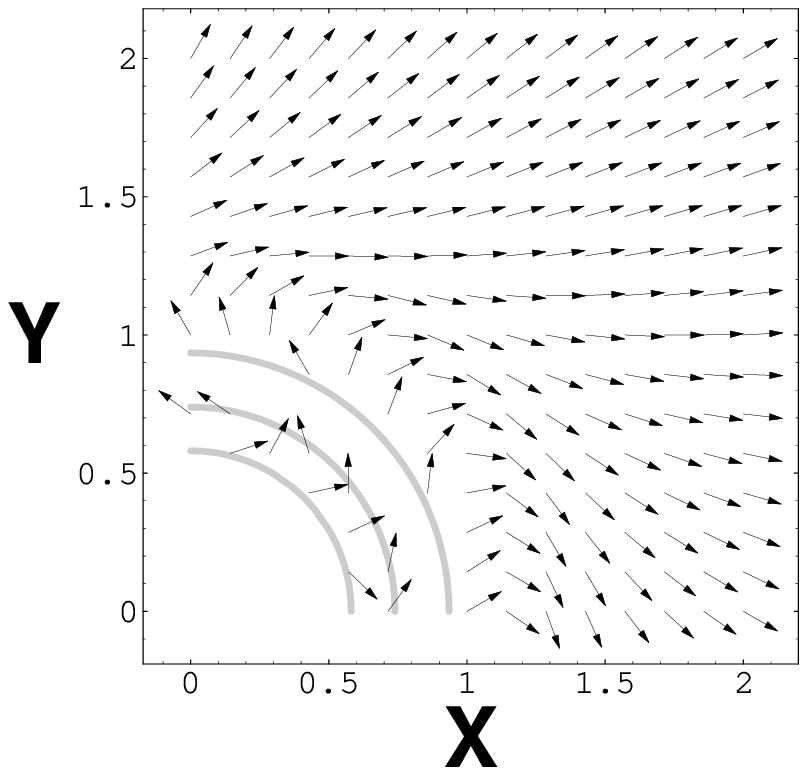}
\includegraphics[width=3.1cm]{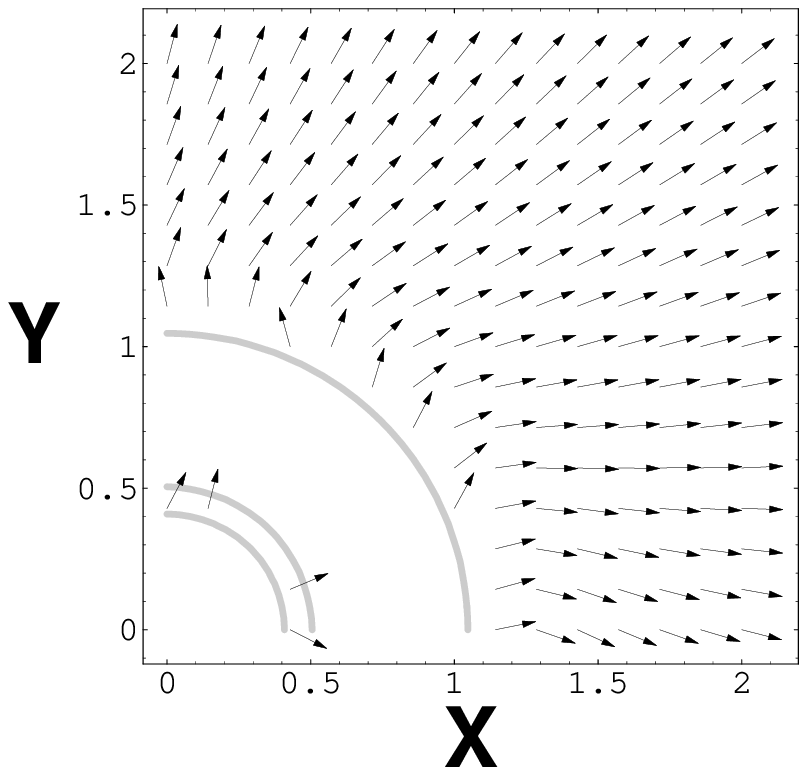}
\includegraphics[width=3.1cm]{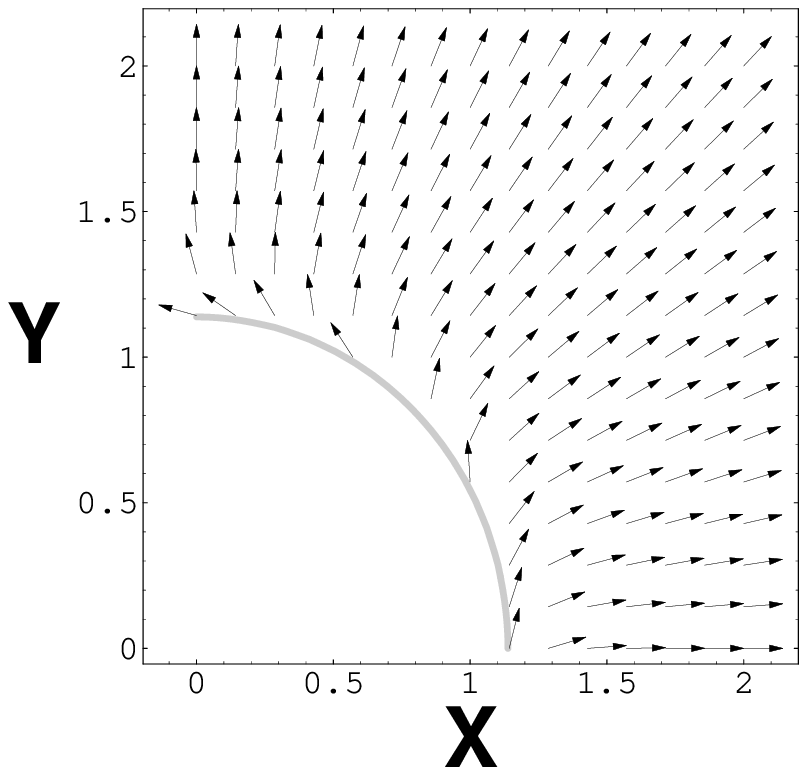}
\includegraphics[width=3.1cm]{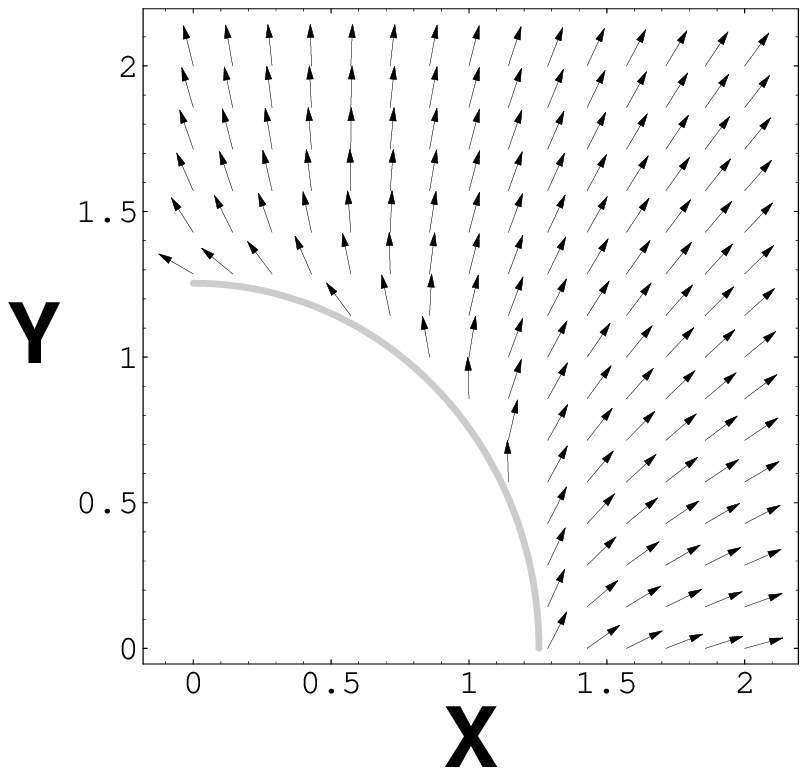}
\includegraphics[width=3.1cm]{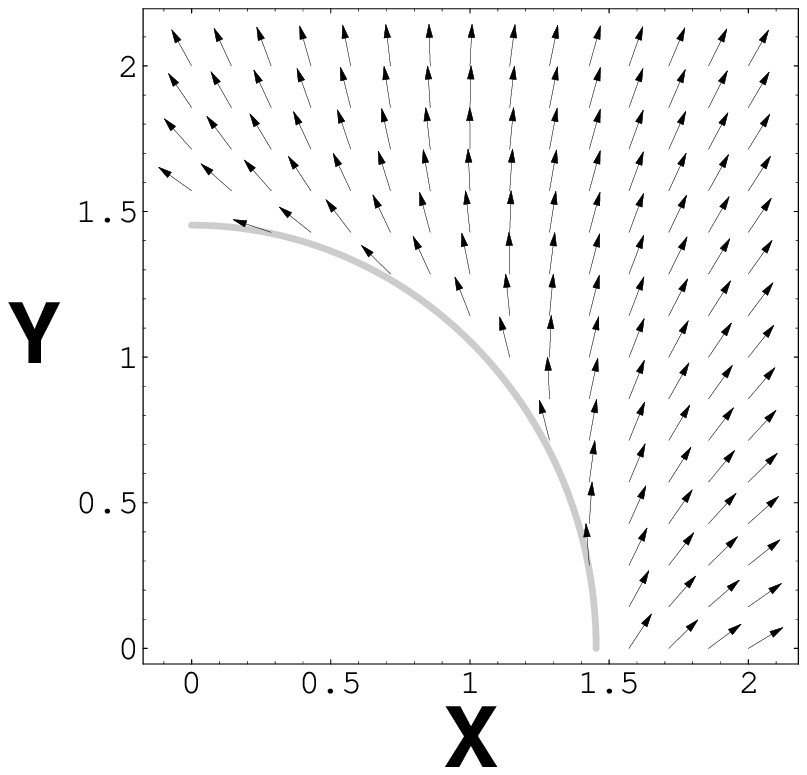}
\caption{The normalized velocity vectors which indicate the flow of evolution
of $X$ and $Y$ (see the text).  (Top row) The commutative case ($\beta=0$).
The figures are plotted for $L=-2, -1, 0, 1, 2$, from left to right.
(Bottom row) The case with the nonconstant noncommutativity.
For the choice of parameters, see the text.
The figures are plotted for $L=-2, -1, 0, 1, 2$, from left to right.
}
\label{fig1}
\end{figure}

We would like to show a concrete example of the solution.
For this purpose, we adopt the following form of $\beta(r)$:
\begin{equation}
\beta(r)=\frac{\beta_0}{1+e^{\mu(r-r_c)}}\,,
\label{beta}
\end{equation}
where $\beta_0$, $\mu$, and $r_c$ are positive constants.%
\footnote{Though the constant $\beta_0$ can take negative values,
we can fix the sign without loss of generality because the equations are invariant
under replacement $\beta(r)\rightarrow-\beta(r)$ and $L\rightarrow-L$.}
This typical form shows that $\beta\sim\beta_0$ for $r<r_c$ and $\beta\sim 0$ for
$r>r_c$, i.e., $\beta$ is finite in the vicinity of the origin of the
minisuperspace, $X=Y=0$. This form mimics a solenoid of radius $r_c$. In
Fig.~\ref{fig1}, we are trying to compare the noncommutative and the
commutative cases. In both cases, we set $\gamma=1$. For the noncommutative case,
we choose
$\beta_0=5$,
$\mu=10$, and $r_c=1$. The evolution of $X=x$ and $Y=y$ can be clearly seen if the
solutions of (\ref{L}) and (\ref{eq2}) are plotted on an $(X, Y)$ plane as shown in
Fig.~\ref{fig1}.
Arrows in the figures indicate the normalized ``velocity'' vectors
\begin{equation}
\frac{1}{\sqrt{\dot{X}^2+\dot{Y}^2}}(\dot{X}, \dot{Y})\,,
\end{equation}
at the points.
In Fig.~\ref{fig1}, the gray curves (quarter circles) indicate that the right-hand
side of (\ref{eq2}) becomes zero.

Roughly speaking, non zero $\beta$ contributes to the ``angular velocity''
$\dot{\phi}$.
Thus, the effective potential
\begin{equation}
V_{eff}(r)=-\frac{1}{2}\gamma^2r^2+\frac{1}{2}\left(\frac{L}{r}
+\frac{r\beta(r)}{2}\right)^2\,,
\label{eff}
\end{equation}
has a positive barrier for a finite $\beta$ even if $L=0$.
For negative $L$ and positive $\beta$, there appears a small region where
$V_{eff}<0$ (in the region of $r<r_c$), so that there appears a region where
the solution is approximately described by (\ref{BX}) and (\ref{BY})
 (with $\Theta=0$ and $B\approx\beta_0$ in this time).
In any case, a large enough $\beta_0$ can forbid the small value of $X$ and
$Y$ simultaneously; In short, $r=\sqrt{X^2+Y^2}<\approx r_c$ is the forbidden
region.

This means that the scale factors of bicosmology  at the beginning of the
classical universe is small but finite, in the sense of ``average'',
$r=\sqrt{X^2+Y^2}\approx r_c$. The discussion on the averaging will be shown later.

\section{noncommutativity in momenta for small scale factors: Quantum system}
\label{sec5}

In this section, we adopt the same system as the one dealt with in the previous
section, and here we consider quantum cosmology
\cite{HH,Hawking,Halliwell,Kiefer0,Kiefer1}.
 The WDW equation is obtained by the following
replacement in the Hamiltonian constraint:
\begin{equation}
\pi_x\rightarrow-i\frac{\partial}{\partial x}\,,\quad
\pi_y\rightarrow-i\frac{\partial}{\partial y}\,.
\end{equation}
Incidentally, one can confirm the commutators
\begin{equation}
[X, \Pi_X]=[Y, \Pi_Y]=i\,,\quad [\Pi_X,
\Pi_Y]=i\left(\beta(r)+\frac{r}{2}\beta'(r)\right)\,,
\end{equation}
by using the set of operators
\begin{equation}
X=x
\,,\quad
Y=y
\,,\quad\Pi_X=-i\frac{\partial}{\partial x}+\frac{\beta(r)}{2}y
\,,\quad
\Pi_Y=-i\frac{\partial}{\partial y}-\frac{\beta(r)}{2}x
\,.
\label{ss}
\end{equation}

The WDW equation $H_{NC}\Psi=0$ in our model,
which is obtained by substituting (\ref{ss}) into (\ref{ncH}),
can be written 
after the coordinate transformation (\ref{ct}) as,
\begin{equation}
\left[
\frac{1}{r}\frac{\partial}{\partial r}r\frac{\partial}{\partial
r}+\frac{1}{r^2}\frac{\partial^2}{\partial \phi^2}-i\beta(r)
\frac{\partial}{\partial \phi}+
\Bigl(\gamma^2-\frac{\beta^2(r)}{4}\Bigr)
r^2\right]\Psi(r,\phi)=0\,,
\end{equation}
where $\Psi$ is the so-called wave function of the universe.
Note that the metric of the minisuperspace is Euclidean (not Lorentzian).

The fundamental solution for this WDW equation takes the form
\begin{equation}
\psi_L(r) e^{-iL\phi}\,,
\end{equation}
and then, the function $\psi_L$ obeys the equation
\begin{equation}
\left[
\frac{1}{r}\frac{\partial}{\partial r}r\frac{\partial}{\partial
r}-\frac{L^2}{r^2}-L\beta(r)+
\Bigl(\gamma^2-\frac{\beta^2(r)}{4}\Bigr)
r^2\right]\psi_L(r)=0\,.
\end{equation}
The general solution then takes the form $\Psi=\sum_{L=-\infty}^\infty
A_L\psi_L(r)e^{-iL\phi}$, where $A_L$ denotes the amplitude. 

For a constant $\beta$, the solution for $\psi_L(r)$ which is finite at $r=0$ is
expressed by using the confluent hypergeometric function as \cite{BS,SB}
\begin{equation}
\psi_L(r)=r^{|L|}e^{
 -i\frac{\sqrt{\gamma^2-\frac{\beta^2}{4}}}{2}r^2
}
{}_1F_1\Biggl(\frac{|L|+1}{2}+\frac{\beta L}{4i\sqrt{\gamma^2-\frac{\beta^2}{4}}};
|L|+1;i\sqrt{\gamma^2-\frac{\beta^2}{4}}r^2\Biggr)\quad(\gamma^2>{\beta^2}/{4})\,,
\end{equation}
\begin{equation}
\psi_L(r)=r^{|L|}e^{
 -\frac{\sqrt{\frac{\beta^2}{4}-\gamma^2}}{2}r^2
}
{}_1F_1\Biggl(\frac{|L|+1}{2}+\frac{\beta L}{4\sqrt{\frac{\beta^2}{4}-\gamma^2}};
|L|+1;\sqrt{\frac{\beta^2}{4}-\gamma^2}r^2\Biggr)\quad(\gamma^2<{\beta^2}/{4})\,.
\end{equation}
Especially, for the commutative case $\beta\equiv 0$,
$\psi_L(r)\propto J_{|L|/2}(\gamma r^2/2)$, where $J_\nu(z)$ is the Bessel
function.

For relatively small values of $\beta$ and finite $L$, the first peak of the
wave function
$\psi_L$ is pushed to a larger $r$ by the ``centrifugal force''.

\begin{figure}[ht]
\centering
\includegraphics[width=4cm]{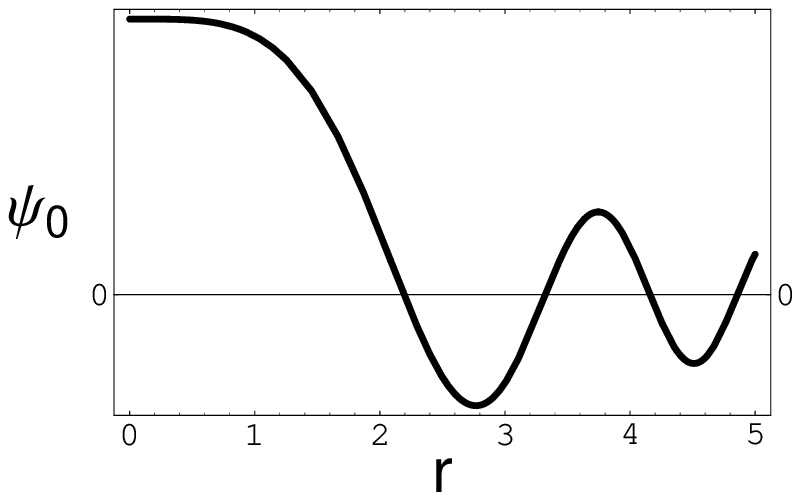}
\includegraphics[width=4cm]{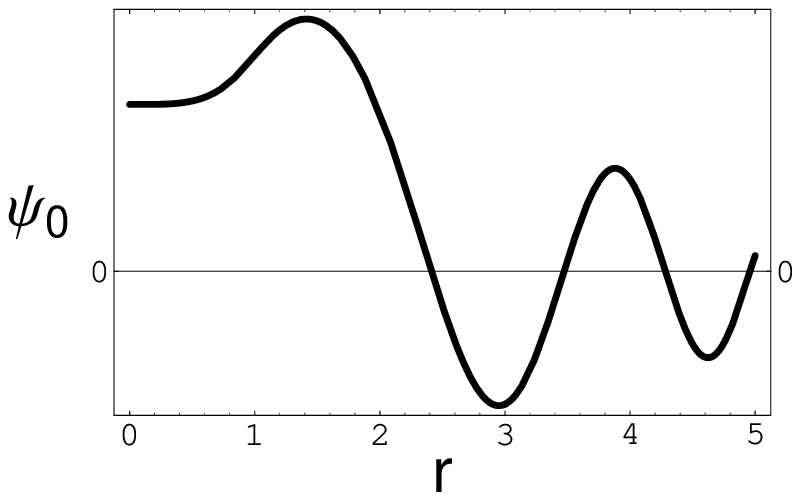}
\includegraphics[width=4cm]{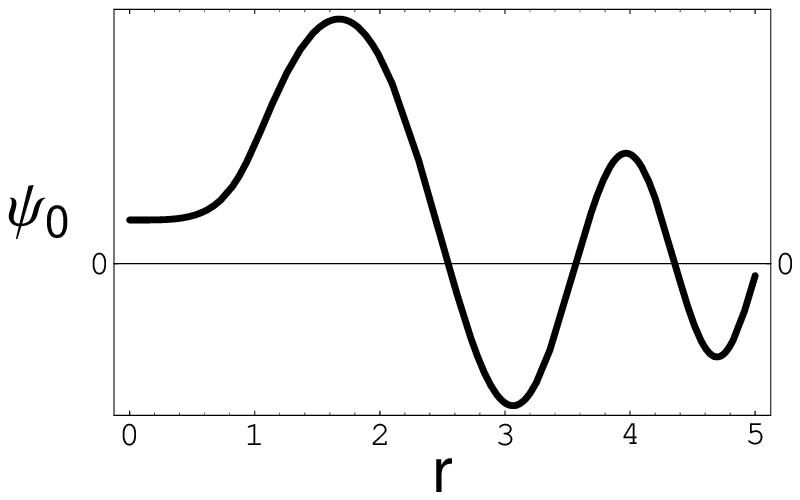}
\includegraphics[width=4cm]{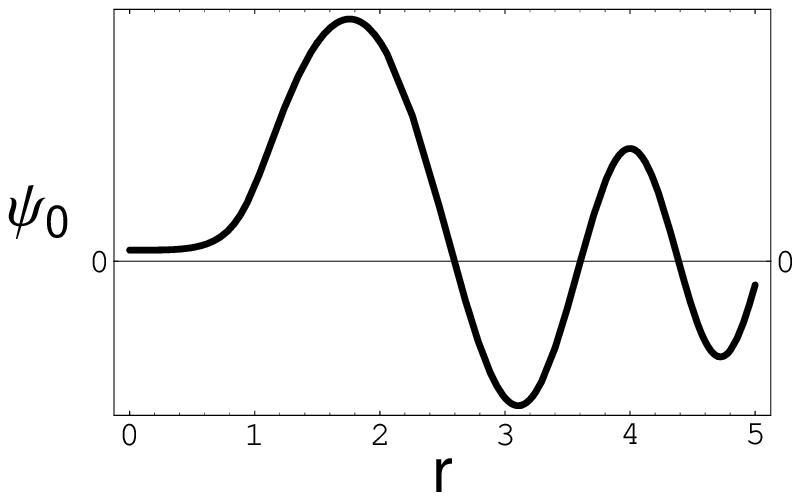}
\caption{
The numerical solutions $\psi_0(r)$ ($L=0$) are plotted for
$|\beta_0|=0, 5, 10, 15$, from left to right.
There is no scale on the vertical axis because an overall coefficient of the
fundamental solution is arbitrary.
}
\label{fig2}
\end{figure}

\begin{figure}[ht]
\centering
\includegraphics[width=7.5cm]{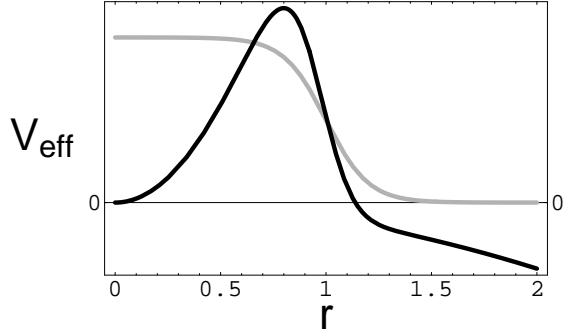}
\caption{A schematic view of the effective potential $V_{eff}(r)$ for $L=0$.
The gray curve indicates $\beta(r)$.}
\label{fig3}
\end{figure}

In the case with the nonconstant noncommutativity for small scale factors, proposed
in the previous section, the situation is the same.
The most interesting case is when $L=0$ with the nonconstant noncommutativity.
In Fig.~\ref{fig2}, we plot $\psi_0(r)$ for various values of $\beta_0$,
where $\beta(r)$ is the same as (\ref{beta}), and $r_c=1$, $\mu=10$, and
$\gamma=1$.
This behavior can be understood by considering the effective potential $V_{eff}$
(\ref{eff}). As seen from Fig.~\ref{fig3}, the noncommutativity, which exists
only in the region where $r$ is small, creates the potential barrier and the first
peak of the wave function is located outside
$r=r_c$; It can be said that the wave function represents the tunneling
phenomenon. The possibility that our universe start with a small but finite scale
(in the present case, $r\approx r_c$) seems to be very interesting;
The scenario is very akin to the model of early quantum cosmology, in which
the potential barrier is supplied by the curvature of space.

Now, we confront the question: why should we focus on the $L=0$ mode?
One answer is the setting of $L=0$ seems most simple and probable.%
\footnote{However, this is not so scientific but rather philosophical argument that
exists universally in study of cosmology.} Another answer is that we might consider
averaging the (``angular'') variables, such that%
\footnote{An additional advantage of the averaging is that we do not have
to worry about the boundary conditions at $x=0$ and at $y=0$.}
\begin{equation}
\frac{1}{2\pi}\int_0^{2\pi}\Psi(r,\phi) d\phi\propto \psi_0(r)\,.
\end{equation}

Even in the classical noncommutative solutions, averaging yielded
interesting results.
Further discussions are provided in the last section.

\section{Conclusion and discussion}
\label{conclusion}
In this work, we investigated classical and quantum solutions for a toy model
of bicosmology.
We considered a symmetric model with two scale factors.
We then derived the classical equations for the model
in the presence of noncommutativities both between the canonical momenta and
between the dynamical variables of the minisuperspace.
Further, we studied the system with the nonconstant noncommutativity in momenta
for small scale factors and explored classical and quantum cosmological solutions.
In this system, the tunneling wave function can be found, which may provide a
new mechanism of the creation of the universe,
even though the curvature of our space is zero.

We must comment on the aforementioned averaging of variables.
According to the original idea put forward in Refs.~\cite{FGMG,FGGM,FGM,MM},
two scale factors represent the causally disconnected regions.
Naturally, it can be supposed that such two regions are extracted from a lot of
typical regions in the universe.
Considering the averaging is therefore not a very new procedure.
However, the average presented here of the scale factors shows strange
fractional powers (such as $x\propto a^{\frac{D-1}{2}}$), so expressing 
a proper averaging in terms of the classical metrics requires
further consideration. When it comes to the wave function of the universe, 
averaging is a speculative but interesting conjecture, because it explains the
creation of a finite-size flat universe.

An important problem is to find the mechanism for determining the critical value
of $r_c$ (as well as the finite $\beta_0$). Unfortunately, this is a difficult
problem with few clues at the moment.

A pressing problem, on the other hand, is to introduce the curvature
 of the space and matter fields into the model.  
We point out that introducing of phantom (scalar) fields
\cite{Caldwell,CKW,DSS,DKS,KS} into the model allows us to choose 
a well-defined boundary condition for
the wave function at the origin of the minisuperspace without changing
the signature of the minisuperspace metric, although there are various
difficulties associated with phantom fields.

In addition, it is worth considering further cases with general noncommutative
configuration variables (such as Sec.~\ref{sec3}) and other possibilities such as
$[X,\Pi_X]=if(\Pi_X,\Pi_Y)$ (see Refs.~\cite{BS,SB} for example),
by classical and quantum approaches.

\appendix

\acknowledgments
We would like to thank Taiga Hasegawa for useful communication.

\bibliographystyle{apsrev4-1}

\end{document}